\documentclass[sigconf]{acmart}
\usepackage[linesnumbered,ruled,vlined]{algorithm2e}
\usepackage{epstopdf}
\usepackage{colortbl}
\usepackage{enumitem}
\usepackage{multirow}
\usepackage[normalem]{ulem}
\useunder{\uline}{\ul}{}
\usepackage{booktabs}
\usepackage{caption}
\usepackage{subcaption}
\usepackage[linesnumbered,ruled,vlined]{algorithm2e}
\usepackage{graphicx}
\usepackage[textsize=small]{todonotes}
\newcommand{\vect}[1]{\boldsymbol{#1}}
\def\header{\vspace{1mm} \noindent}

\definecolor{LightSteelBlue}{RGB}{213,229,255}

\AtBeginDocument{%
  \providecommand\BibTeX{{%
    \normalfont B\kern-0.5em{\scshape i\kern-0.25em b}\kern-0.8em\TeX}}}

\copyrightyear{2025} 
\acmYear{2025} 
\setcopyright{cc}
\setcctype{by}
\acmConference[WWW '25]{Proceedings of the ACM Web Conference 2025}{April 28-May 2, 2025}{Sydney, NSW, Australia}
\acmBooktitle{Proceedings of the ACM Web Conference 2025 (WWW '25), April 28-May 2, 2025, Sydney, NSW, Australia}
\acmDOI{10.1145/3696410.3714651}
\acmISBN{979-8-4007-1274-6/25/04}

\begin{document}

\title{Rumor Detection on Social Media with Reinforcement Learning-based Key Propagation Graph Generator}

\author{Yusong Zhang}
\authornote{Yusong Zhang and Kun Xie are joint first authors.}
\affiliation{%
  \institution{The Chinese University of Hong Kong}
  \country{Hong Kong}
  }
\email{1155107841@link.cuhk.edu.hk}

\author{Kun Xie}
\authornotemark[1]
\affiliation{%
  \institution{The Chinese University of Hong Kong}
  \country{Hong Kong}
  }
\email{xiekun@se.cuhk.edu.hk}

\author{Xingyi Zhang}
\affiliation{%
  \institution{Mohamed bin Zayed University of Artificial Intelligence}
  \country{United Arab Emirates}
  }
\email{xingyi.zhang@mbzuai.ac.ae}

\author{Xiangyu Dong}
\affiliation{%
  \institution{The Chinese University of Hong Kong}
  \country{Hong Kong}
  }
\email{xydong@se.cuhk.edu.hk}

\author{Sibo Wang}
\authornote{Sibo Wang is the corresponding author.}
\affiliation{%
  \institution{The Chinese University of Hong Kong}
  \country{Hong Kong}
  }
\email{swang@se.cuhk.edu.hk}

\renewcommand{\shortauthors}{Yusong Zhang, Kun Xie, Xingyi Zhang, Xiangyu Dong, and Sibo Wang}

\begin{abstract}
The spread of rumors on social media, particularly during significant events like the US elections and the COVID-19 pandemic, poses a serious threat to social stability and public health. Current rumor detection methods primarily rely on propagation graphs to improve the model performance. However, the effectiveness of these methods is often compromised by noisy and irrelevant structures in the propagation process. To tackle this issue, techniques such as weight adjustment and data augmentation have been proposed. However, they depend heavily on rich original propagation structures, limiting their effectiveness in handling rumors that lack sufficient propagation information, especially in the early stages of dissemination. In this work, we introduce the {\em \underline{K}ey \underline{P}ropagation Graph \underline{G}enerator (KPG)}, a novel reinforcement learning-based framework, that generates contextually coherent and informative propagation patterns for events with insufficient topology information and identifies significant substructures in events with redundant and noisy propagation structures. KPG comprises two key components: the {\em \underline{C}andidate \underline{R}esponse \underline{G}enerator (CRG)} and the {\em \underline{E}nding \underline{N}ode \underline{S}elector (ENS)}. CRG learns latent variable distributions from refined propagation patterns to eliminate noise and generate new candidates for ENS, while ENS identifies the most influential substructures in propagation graphs and provides training data for CRG. Furthermore, we develop an end-to-end framework that utilizes rewards derived from a pre-trained graph neural network to guide the training process. The resulting key propagation graphs are then employed in downstream rumor detection tasks. Extensive experiments conducted on four datasets demonstrate that KPG outperforms current state-of-the-art methods.

\end{abstract}



\maketitle

\section{Introduction}
\label{sec:intro}
\begin{figure}
    \centering
    \includegraphics[width=80mm]{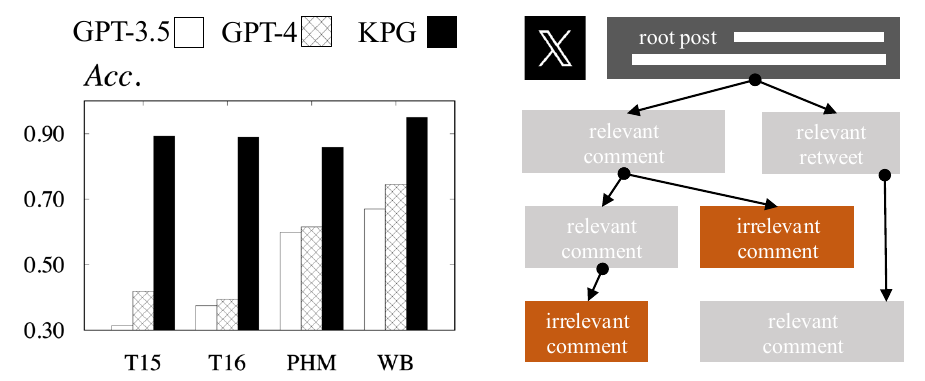}
   \vspace{-3mm}
    \caption{Left: classification accuracy on four datasets. Right: an example propagation graph on the social network. }
    \label{fig:llm-prop}
  \vspace{-3ex}
\end{figure}

As a primary medium for information dissemination, social media eliminates temporal and spatial constraints on communication, facilitating the spread of information. However, the popularity of certain topics on social media often leads to the propagation of intentional or unintentional misinformation \cite{zhou20fakenewssurv}, negatively impacting individual health and social stability \cite{YangSWG0019, zhou20recovery}. Unfortunately, individuals who encounter unverified information often struggle to distinguish between truth and falsehood. This difficulty can inadvertently lead to the further spread of misinformation through social media platforms~\cite{Rubin10, zhou20fakenewssurv} and word-of-mouth effect~\cite{feng2024efficient, guo2023efficient, guo2022influence}.
Additionally, this challenge extends to sophisticated tools like large language models, which also struggle to accurately identify rumors on social networks, as shown in the left subfigure of Fig. \ref{fig:llm-prop}. The complexity of this issue underscores the need for more nuanced approaches to misinformation detection and management.

In recent years, rumor detection has gradually shifted from traditional machine learning approaches \cite{wu15rwkernel,ma17twitter1516,zhou19pattern,rosenfeld20wlkernel}, which require the manual definition of content, structural, or information source features, to deep learning approaches \cite{ma18rvnn,Khoo20,WeiXM19,LiZS19,wang18eann,ZhangLLY19,MaHGCZZ22}. 
Notably, {\em graph neural networks (GNNs)}~\cite{kipf2016gcn, velivckovic2017gat, dong2023rayleigh,dong35smoothgnn,zhang2024ficom} effectively combine node textual features and graph propagation topology, achieving superb detection performance~\cite{bian2020bigcn, wei2021ebgcn, he21rdea, sun22gacl, zhang23dcerd}.
As illustrated in the right subfigure of Fig. \ref{fig:llm-prop}, root nodes in the propagation graphs represent the original posts, while other nodes represent comments or retweets received during the spread. Edges indicate the relationships of commenting and retweeting between nodes. 

Although using propagation information improves rumor detection accuracy, irrelevant or untrustworthy comments on social networks significantly impair the reliability of propagation graphs in identifying the veracity of claims, as noted in \cite{wei2021ebgcn,sun22gacl}. 
To address this challenge, techniques such as weight adjustment \cite{wei2021ebgcn, wei2022fgcn} and data augmentation \cite{he21rdea, sun22gacl, liu2023trustrd} have been introduced. However, these strategies rely on graphs with abundant structural information, which is often lacking in rumors, especially in the early stages of their spread on real-world social networks. 
Specifically, in the early stages of spread, interactions among social network users are limited due to the short duration of engagement. The original post has not yet garnered a significant number of comments or retweets, resulting in a relatively small propagation graph. This small size renders existing solutions ineffective because they are based on edge reweighting or graph augmentation, which adjust the topology of the input propagation graph but cannot generate or expand the graph itself.
Furthermore, as noted in recent studies \cite{Liu22longtailgraph, Zhao22topolmb}, existing graph neural networks used in rumor detection methods struggle to effectively address events with insufficient topological information. The absence of distinguishable structures in small graphs is the primary reason for unsatisfactory results \cite{Liu22longtailgraph}. Overall, designing an effective rumor detector that can handle both noisy propagation graphs with irrelevant comments and small propagation graphs in the early stage of spread remains an unresolved problem.

Motivated by these challenges, we propose KPG (\underline{K}ey \underline{P}ropagation Graph \underline{G}enerator) to generate enhanced key propagation graphs for rumor detection. Our KPG consists of two key components: the {\em Candidate Response Generator (CRG)} and the {\em Ending Node Selector (ENS)}. First, we propose the CRG module to address situations where propagation graphs are too small to provide sufficient candidates for selection. CRG generates new responses that are aligned with the structure and contextual information of the graph. These responses are derived from realistic yet refined noise propagation patterns, aiming to enhance the discrimination potential of the augmented graphs. Next, we introduce ENS to explore the candidate graph from both local and global perspectives. It identifies the key propagation graph by iteratively selecting critical nodes from the candidate set. Instead of choosing nodes based on the semantic relevance of comments, ENS evaluates each node based on the contribution of its feature and topology information to the classification performance of the event. By incorporating these two modules, our KPG can select distinguishable key patterns from the original noisy propagation graphs and expand small propagation graphs with informative new comments, thereby improving the overall model performance.

We further incorporate a {\em reinforcement learning (RL)} framework to integrate the candidate node generation and key node selection processes with graph classification accuracy. To achieve this, we design rewards based on the classification accuracy obtained from a pre-trained {\em Graph Neural Network (GNN)} classifier. Specifically, we offer rewards for newly generated responsive features that show superior performance compared to the original features in terms of classification accuracy. Additionally, we encourage the selection of nodes that enhance the discriminative capability of the graph and maximize the expected cumulative performance improvement. Finally, we develop KPG as an end-to-end framework that trains the ENS and CRG alternately. This allows both components to generate intermediate results for each other and be optimized iteratively. On one hand, the CRG generates new responses to supplement the candidate set of the ENS, enabling ENS to overcome limitations imposed by insufficient input information. On the other hand, the key patterns identified by ENS serve as training samples for CRG, facilitating the extraction of latent variable distributions from realistic and distinguishable propagation patterns. The final key propagation graphs are then utilized to train a downstream GNN classifier for the rumor detection task.

In summary, our contributions are as follows:
\begin{itemize}[topsep=0mm, partopsep=0pt, itemsep=0pt, leftmargin=10pt]
    \item We propose KPG, a novel rumor detection model comprising two interdependent modules. It can augment small propagation graphs with contextually relevant, reliable responses and precisely select the indicative key nodes considering both feature and structural information.
    \item We incorporate the RL framework into propagation graph-based rumor detection. The carefully designed rewards improve discriminability during key graph generation process.
    \item Extensive experiments show that KPG achieves state-of-the-art performance in the rumor detection task.
\end{itemize}

\section{Preliminaries}
\subsection{Notation} 

Let $s=\left\{r, G=(V, E, \vect{X})\right\}$ denote an event, where $r$ is the source post and $G$ indicates the propagation graph of the source post. $G$ is a directed acyclic tree rooted at the source post $r$. The node set $V=\{r, v_1, v_2, \cdots, v_{n_s-1}\}$ contains all $n_s$ comments and retweets in the propagation process. The text stored in node $u$ is denoted as $\text{text}_u$. The edge set $E=\left\{(v_i, v_j)|v_i, v_j \in V\right\}$ represents the propagating relation between posts, i.e., there exists an directed edge $(v_i, v_j) \in E$ if $v_j$ is a comment or retweet of $v_i$. $\vect{X}\in\mathbb{R}^{n_s \times d}$ contains the initial features of posts in the event $s$. Each row of the feature matrix $\vect{X}[v_i]$ is a $d$-dimensional vector of the corresponding post $v_i$. Table \ref{tab:notation} in Appendix \ref{sec:notation} lists the frequently used notations in this paper.

\noindent
{\bf Rumor Detection.}
Let $y_s\in\mathcal{Y}$ be the label of the event $s$, indicating the veracity of the event. The label set $\mathcal{Y}$ consists of non-rumors, false rumors, true rumors, and unverified rumors. Some datasets contain only two classes: rumors and non-rumors. Given a set of events $\{s_0, s_1, \cdots\}$ obtained from social networks, the goal of rumor detection is to predict the veracity of each event.

\noindent
{\bf Key Propagation Generation}
We formulate the key propagation generation problem within the reinforcement learning framework. 
At step $t$, the {\em state} is the key propagation graph $g_t=\{V_{g_t}, E_{g_t}, \vect{X}_{g_t}\}$, where $|V_{g_t}| = n_t$ and $|E_{g_t}| = m_t$. The initial state $g_0$ contains only the root node $r$ and its feature $\vect{X}[r]$.
The {\em action} $a_t=(v_t, e_t)$ represents the node and edge to be added to $g_t$ for generating $g_{t+1}$.

\subsection{Related Work}
\label{sec:related_work1}
We review existing \textit{propagation graph-based rumor detection models} in this section, deferring the discussion of other related work to Appendix \ref{sec:related-work2}. Specifically, BiGCN~\cite{bian2020bigcn} pioneered the application of GCN to enhance the exploitation of propagation graphs through both top-down and bottom-up directions. EBGCN~\cite{wei2021ebgcn} and FGCN~\cite{wei2022fgcn} proposed edge weight adjustment mechanisms to capture the interaction intensity between comments. Subsequently, models including RDEA~\cite{he21rdea}, GACL~\cite{sun22gacl}, TrustRD~\cite{liu2023trustrd}, and others~\cite{MaHGCZZ22,9892019, cui2024ragcl} incorporated data augmentation strategies such as edge perturbation and masking to enhance model robustness through contrastive learning.
Furthermore, researchers have introduced innovative techniques to improve the accuracy and reliability of rumor detection, including position-aware adversarial response generation~\cite{song21aard}, diverse counterfactual evidence exploitation~\cite{zhang23dcerd}, and semantic evolvement extraction~\cite{tao24gard}.
However, existing approaches heavily rely on original topology, limiting their effectiveness on small-scale graphs with insufficient structural information. In contrast, our KPG incorporates two interdependent modules that generate key patterns for propagation graphs of varying sizes, demonstrating superb performance in experimental evaluations.

\section{Key Propagation Graph Generator}
\label{sec:solution}

In this section, we present our KPG model, which comprises two key components: the {\em Candidate Response Generator (CRG)} and the {\em End Node Selector (ENS)}. To address the limitations of existing models in handling small input graphs with insufficient topological information, we propose the CRG module to generate realistic and reliable responses for expanding the propagation graph. Subsequently, the ENS module generates a probability distribution for each node, indicating the likelihood of the node being selected during the step-by-step construction of the key propagation graph. Specifically, we select propagation patterns that are critical and indicative for classification, while filtering out irrelevant and noisy comments. An RL framework is utilized to design rewards for both modules, further enhancing the model performance on the generated key propagation graphs. With the guidance of carefully designed rewards, the classification accuracy of the key propagation graph in the current state is encouraged to surpass that of its preceding state, ensuring continuous improvement throughout the generation process. The rest of this section is arranged as follows. First, we present the overall framework of KPG in Section \ref{sec:overview}. Then, we elaborate on CRG and ENS in Sections \ref{sec:crg} and \ref{sec:ens}, respectively. Finally, we illustrate the reward functions used for guiding CRG and ENS, and present the learning algorithm in Section \ref{sec:training}.

\begin{figure}[t]
    \centering
    \begin{small}
    \includegraphics[width=80mm]{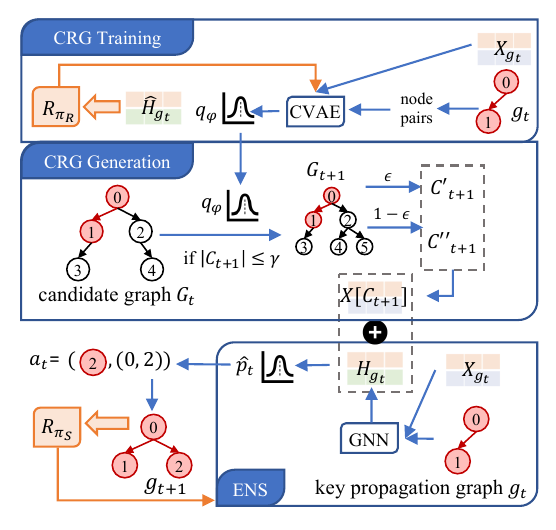} 
    \vspace{-3mm}
    \caption{Framework of KPG.}
    \label{fig:kpg}
    \end{small}
    \vspace{-3mm}
\end{figure}

\subsection{Overview} 
\label{sec:overview}

Figure \ref{fig:kpg} illustrates the KPG framework at step $t$, showing the generation process of the key propagation graph. At the bottom right of Figure \ref{fig:kpg}, a toy example of a key propagation graph $g_t$ with two nodes at step $t$ is presented. Suppose the current candidate graph $G_t$ is the graph at the middle left of Figure \ref{fig:kpg}. If the number of candidates is less than $\gamma$, the CRG $\pi_R(\cdot)$ generates new responses drawn from a learned latent distribution. In this example, node $5$ is added into the new candidate graph $G_{t+1}$, aiding in generating the candidate set $C_{t+1}$. Next, the ENS $\pi_S(\cdot)$ encodes the current graph $g_t$ using a GNN and concatenates the node representations with candidate features $X[C_{t+1}]$. An MLP is then utilized to derive the probability of each available action $a_t$. For example, at the bottom left of Figure \ref{fig:kpg}, graph $g_{t+1}$ is updated based on action $a_t =(2, (0,2))$. Finally, the rewards $R_{\pi_S}$ and $R_{\pi_R}$ are designed to increase the discriminative capacity of the generated key graph. Next, we elaborate on the CRG component in KPG.

\subsection{Candidate Response Generator} 
\label{sec:crg}
In this section, we present the CRG component $\pi_R(\cdot)$.
Existing data augmentation approaches primarily focus on perturbing the input structure. However, perturbation-based models struggle with small propagation graphs, which inherently lack sufficient topology and feature information. In contrast, CRG presents a novel approach by generating entirely new content as responses to the given nodes, thereby augmenting the graph in a contextually coherent manner. Specifically, CRG supports event classification from three aspects: 
\begin{itemize}[topsep=0pt,leftmargin=1.1em]
    \item {\em (i)} Contextually coherent: generating responses relevant to the original context in the propagation graph.
    \item {\em (ii)} Realistic yet refined from noise: generating responses that reflect real-world social network propagation patterns while being robust to noisy and irrelevant comments.
    \item {\em (iii)} Discriminative: generating distinct responses that contribute to rumor detection and downstream performance improvement.
\end{itemize}
Among these objectives, we achieve {\em (iii)} through carefully designed rewards, discussed in Section \ref{sec:training}. The mechanisms behind {\em (i)} and {\em (ii)} are detailed in this section. 

To achieve objective {\em (i)}, we construct a {\em conditional variational auto-encoder (CVAE)} \cite{Sohn2015cvae}. This extension of the traditional VAE \cite{kingma14vae} incorporates condition variables to generate new responses following a latent distribution. By extracting this distribution from real-world propagation patterns, CVAE effectively captures the data distribution of existing responses, thereby facilitating the generation of responses that maintain contextual coherence.

To achieve objective {\em (ii)}, we leverage the currently identified key propagation graph $g_t$ rather than the original input graphs as training data for CRG, filtering out noisy and irrelevant comments. This approach allows CRG to learn from propagation patterns that significantly contribute to the classification task, alleviating the impact of noisy comments.

\header{\bf CRG Training.}
Let $g_t$ be a key propagation graph at step $t$. We extract all pairs from the current key propagation graph $g_t$ as the training data. Let $(u, v) \in g_t$ represent a context-response node pair, where $u$ denotes the {\em context} and $v$ denotes the {\em response}. Initially, we employ a GRU \cite{cho14gru} to encode the text of $u$ and $v$ into representations $\vect{h}_u$ and $\vect{h}_v$, respectively. To address the one-to-many problem~\cite{Sohn2015cvae}, we concatenate the latent representation of $u$ with its source post $r$ to strengthen the connection between $\vect{h}_u$ and its central topic:
\begin{align}
    & \vect{h_u} = \text{CONCAT} \left( \text{GRU}( \text{text}_u), \text{GRU} (\text{text}_r ) \right), \label{eq:h_u} \\ 
    & \vect{h_v} = \text{GRU}\left(\text{text}_v\right). \nonumber    
\end{align}
Next, representations $\vect{h}_u$ and $\vect{h}_v$ are fed into the CVAE model, which includes an encoder $q_\varphi(\vect{z}|\vect{h}_u, \vect{h}_v)$ with latent space representation $\vect{z}$, an MLP decoder, and a GRU decoder. Assume that $\vect{z}$ follows a Gaussian distribution $\mathcal{N}(\vect{\mu}, \vect{\sigma}^2)$. During the encoding phase, $\vect{h}_u$ and $\vect{h}_v$ are concatenated and fed into an MLP to learn distribution parameters of the encoder: 
$$
    \vect{\mu} = \text{MLP}_\mu \left( \text{CONCAT} (\vect{h}_u, \vect{h}_v) \right), \quad
    \vect{\sigma}^2 = \text{MLP}_\sigma \left( \text{CONCAT} (\vect{h}_u, \vect{h}_v) \right).
$$
A sample $\vect{z}' \sim q_\varphi(\vect{z} \vert \vect{h}_u, \vect{h}_v)$ is then drawn from the learned distribution $\mathcal{N}(\vect{\mu}, \vect{\sigma}^2)$ using the reparameterization trick \cite{Sohn2015cvae}. This sample is then combined with the context representation $\vect{h_u}$ to reconstruct the response through the MLP decoder:  
\begin{align}
    \hat{\vect{h}_v} = \text{MLP}_{dec}( \text{CONCAT} (\vect{z}', \vect{h}_u) ). 
    \label{eq:h_v_hat}
\end{align}
The goal of CRG is to minimize the reconstruction error between representations of the generated response $\hat{\vect{h}}_v$ and the original response $\vect{h}_v$, thereby extracting contextually coherent distributions. Detailed loss functions and rewards are presented in Section~\ref{sec:training}. Next, we elaborate on the candidate set generation process after the completion of the CRG training.

\noindent
{\bf CRG Generation.}
In the generation phase of CRG, we set a threshold $\gamma$ for the size of the candidate set, which is initialized using the original graph. If the number of candidate nodes exceeds $\gamma$, the candidate graph remains unchanged and is inherited in the next step. Otherwise, we uniformly sample nodes from the current candidate graph $G_t$ as contexts and invoke the trained CRG to generate new response nodes for these selected context nodes until the candidate size requirement is satisfied. 

Suppose at step $t$, the size of the current candidate set is less than or equal to $\gamma$, and we select a context node $i$ from the candidate graph $G_t$. We first generate the topic-aware representation $\vect{h}_i$ based on Eq. ~\eqref{eq:h_u}. Next, $\vect{z}'$ is sampled following the learned distribution $q_\varphi$. We then combine $\vect{z}'$ and $\vect{h}_i$ to generate the response $\vect{h}_j$ according to Eq.~\eqref{eq:h_v_hat}. The text of node $i$ can be generated using the GRU decoder based on the decoded representation $\vect{h}_j$. After that, we add the node $j$ and edge $(i,j)$ to the candidate graph $G_t$. This process will continue until the size of candidate set exceeds $\gamma$.

After obtaining the updated candidate graph $G_{t+1}$ with sufficient nodes, we select candidates for the current identified key propagation graph $g_t$ using a hybrid selection method. Specifically, we consider two types of candidates. The first type, referred to as {\em local candidates}, consists of the nodes that are directly connected to nodes in the current key propagation graph $g_t$: 
$$
    C^{'}_{t+1}=\{v\in {V}_{G_{t+1}}|(u, v) \in {E}_{G_{t+1}},  u\in V_{g_t}\}.
$$
The second type of nodes comprises those that exist in the candidate graph ${G}_{t+1}$ but are not present in the current propagation graph $g_t$. These nodes are referred to as {\em global candidates}:
$$
    C^{''}_{t+1}=\{v \in {V}_{G_{t+1}}|v \notin V_{g_t}\}.
$$

\noindent
{\bf Example.} In the example illustrated in Figure \ref{fig:kpg}, $G_{t+1}$ consists of 6 nodes, and the current key propagation graph $g_t$ contains node 0 and node 1. Therefore, the local candidate set $C_{t+1}^{'} = \{2,3\}$ and the global candidate set $C_{t+1}^{''} = \{2,3,4,5\}$.

By introducing the second type of nodes, we incorporate a restart mechanism that allows all unselected nodes to be considered during the node selection process. This approach strikes a balance between localized exploration around the existing key nodes and a global search across the entire candidate graph. Besides, a trade-off parameter $\epsilon\in [0, 1]$ is introduced to control the balance between these two types of candidates. At each step $t$, we either set $C_{t+1}=C^{'}_{t+1}$ with a probability of $\epsilon$ to locally explore the boundary of the current key propagation graph $g_t$, or set $C_{t+1}=C^{''}_{t+1}$ with a probability of $(1-\epsilon)$ to restart the exploration from one of the other nodes that have not been selected before, enabling a global search across of the entire graph. In our experiments, we set $\epsilon = 0.8$ by default.

\subsection{Ending Node Selector} 
\label{sec:ens}
In this section, we introduce the Ending Node Selector $\pi_S(\cdot)$. At each step $t$, we first adopt CRG to update the candidate graph from ${G}_t$ to ${G}_{t+1}$. Subsequently, based on the current key propagation graph $g_t$ and the candidate graph ${G}_{t+1}$, ENS predicts the probabilities of candidates being selected for action $a_t$. These probabilities are determined by the potential improvement in event classification performance. ENS aims to select the node that contributes most significantly to enhancing classification accuracy while disregarding those that are irrelevant to the veracity of the event.

To achieve this, we assess each node in the candidate set $C_{t+1}$ based on its own features and its connection with the current key graph $g_t$. Specifically, ENS first encodes the current key propagation graph $g_t$ using a GCN. Next, it augments the features of candidate nodes through concatenation operations. The augmented features of candidates are then fed into an MLP, followed by a softmax function, to generate the probability distribution for the next action $a_t$. Finally, the key propagation graph is updated accordingly.

\header
{\bf ENS Component.}
We aggregate neighbor features from the graph structure using a {\em Graph Convolutional Network (GCN)} \cite{kipf2016gcn}, such that both textual and topological information in the current key graph can be incorporated. Let $\hat{\vect{A}}_{g_t}$ is the normalized adjacency matrix of $g_t$ with self-loops and $\vect{X}_{g_t}$ be the feature matrix associated with nodes in $g_t$, we derive the representation for nodes in $g_t$ through a two-layer GCN model:
\begin{align*}
    \vect{H}_{g_t}= \delta( \hat{\vect{A}}_{g_t} \delta (\hat{\vect{A}}_{g_t} \vect{X}_{g_t} W_1) W_2), 
\end{align*}
where $\delta(\cdot)$ is the activation function, and $W_1$ and $W_2$ are learnable parameter matrices.

Next, we augment the candidate features by concatenating the input candidate features with the representations of their parents in $\vect{H}_{g_t}$. Given a candidate node $v \in C_{t+1}$ with input feature $\vect{X}[v]$, if its parent node $v_p \in G_{t+1}$ also exists in the current propagation graph $g_t$, ENS directly concatenates its feature $\vect{X}[v]$ with the representation of $v_p$: 
$$
    {\vect{X}}'[v] = \text{CONCAT}({\vect{X}}[v], \vect{H}_{g_t}[v_p]).
$$
Otherwise, if the parent node $v_p \notin V_{g_t}$, we pad its feature with a zero vector:
$$
    {\vect{X}}'[v] = \text{CONCAT}({\vect{X}}[v], \vect{0}).
$$

After that, we employ an MLP with a softmax function to predict the probability of each candidate being selected. Let $\vect{X}'[C_{t+1}]$ be the collection of augmented features of candidates in set $C_{t+1}$. Then, the probability distribution $\vect{p}_t$ is computed as follows:
$$
    \vect{p}_t = \text{softmax} \left(\text{MLP} \left( {\vect{X}}'[C_{t+1}] \right) \right).
$$

Finally, based on the probability distribution $\vect{p}_t$, action $a_t$ selects a new node $v_t$ from $C_{t+1}$ to be added into $g_t$. The probability of node $v_t$ being selected by action $a_t$, denoted as $\Pr[a_t=(v_t, e(v_t))]$, is determined by the distribution $\vect{p_t}$:
$$
    \Pr[a_t=(v_t, e(v_t))]=\vect{p}_t [v_t], \quad \forall v_t \in C_{t+1}.
$$
Here, $e(v_t)$ represents the edge corresponding to $v_t$. To explain, if the parent node $v_p$ of $v_t$ exists in $g_t$, we set $e(v_t)=(v_p, v_t)$; otherwise, we set $e(v_t)=(r, v_t)$ to emphasize the importance of the root node $r$. The key propagation graph $g_t$ is then updated to $g_{t+1}$ by adding the newly selected node and edge in action $a_t$: 
$$
    g_{t+1}= \left( V_{g_t}\cup v_t, E_{g_t} \cup e(v_t), \vect{X}_{g_t}\cup {\vect{X}}[v_t] \right).
$$

\noindent
{\bf Remark.}
As discussed in Section \ref{sec:crg}, when generating the candidate set $C_{t+1}$, CRG considers two types of candidates and introduces a trade-off parameter $\epsilon$. This strategy is similar to the restart probability in the random walk \cite{Tong06rwr, zhang2024towards, hou2023personalized, hou2021massively,fora,foratods, PAFO, treesvd, HubPPR}. We either select the next nodes locally from the out-neighbors of the currently selected nodes with $\epsilon$ probability, or restart from another node in the remaining part of the graph $G_{t+1}$ with $(1-\epsilon)$ probability, enabling a global exploration. It is important to note that when restarting from other nodes, it is possible to select a node $v_{t+1}$ that is not directly connected to the current key propagation graph $g_t$. In such cases, we add an edge between the root node $r$ and $v_{t+1}$. This allows CRG to explore additional key propagation patterns without being limited by the structure of the current graph.

\subsection{Model Integration} 
\label{sec:training}
In this section, we introduce the rewards designed to guide the training of both the ENS and CRG modules. The objective is to select actions that maximize the expected improvement in classification accuracy during the key propagation graph generation process. 

\header
\textbf{CRG Reward.} 
Recap from Section \ref{sec:crg} that the CRG aims to generate coherent responses that align with the contextual information in the graph, thereby fulfilling objectives {\em (i)} and {\em (ii)}. Besides, CRG seeks to select informative responses in the propagation graph that improve the overall classification performance, which corresponds to objective {\em (iii)}. To achieve this, we design the reward according to the improvement of classification accuracy obtained by the updated features of responses. Let $y_s$ denote the ground-truth label of event $s$. The reward for CRG is defined as follows: 
\begin{equation}
\label{eq:reward-crg}
    R_{\pi_R}^{(t)}=e^{-(f({g^{'}_t})[y_s]-f({g_t})[y_s])},
\end{equation}
where $f(\cdot)$ is a BiGCN classifier with primary discrimination ability.  
$g_t$ is the current key propagation graph with original features, and $g^{'}_t$ is the graph that shares the same topology as $g_t$ but is associated with updated features generated by CRG. If CRG successfully extracts informative features during training, i.e., the updated features lead to higher classification accuracy than the original features, we reward the model; otherwise, we impose a penalty. Specifically, if the generated responses contribute to the improved classification confidence, we have $R^{(t)}_{\pi_R}<1$, encouraging the module to maintain its well-trained state. In contrast, if the responses are not informative, we obtain $R^{(t)}_{\pi_R}\geq 1$, pushing the module away from its current unsatisfactory state. 

Besides, we adopt the stochastic gradient variational Bayes framework to optimize the reconstruction error and the KL divergence between the variational distribution and the prior distribution. The loss function of CRG is defined as follows:
\begin{equation}
\label{eq:crg-loss}
    \begin{aligned}
        \mathcal{L}_{\pi_R}^{(t)} = R^{(t)}_{\pi_R}\cdot \Sigma ( & \mathbb{E}_{q_{\phi}(\vect{z}|\vect{h_u},\vect{h_v})} [\log p_{\theta}(\vect{h_v}|\vect{z},\vect{h_u})]\\
        & -KL(q_{\phi}(\vect{z}|\vect{h_u},\vect{h_v})\Vert p_{\theta}(\vect{z}|\vect{h_u})) ).
    \end{aligned}
\end{equation}

\header
\textbf{ENS Reward.}
The ENS reward aims to generate a more discriminative key propagation graph after adding a new node. To achieve this, we design the reward of ENS from two aspects: {\em (i)} the improvement of the prediction score and {\em (ii)} the enhancement of future performance estimation. Specifically, we aim to achieve a higher prediction score for class $y_s$ derived from the classifier $f(\cdot)$ on the updated key graph $g_{t+1}$ compared to $g_t$. Simultaneously, we aim to maximize the future performance of $g_{t+1}$ in the subsequent steps. 

To simulate future actions, we employ a modified Rollout \cite{Bertsekas99rollout} to select up to $(l-1)$ additional nodes from the current candidate set. These $l$ nodes, including the selected one, are incrementally added to the current key propagation graph $g_t$, thereby simulating potential actions over the next $l$ steps. We then combine the current and the future prediction scores to calculate an overall score for the updated key propagation graph $g_{t+1}$:
$$
    r_{{t+1}} =\frac{1}{2}\left( f(g_{t+1})[y_s] + \frac{1}{l}\left( \sum^{l}_{i=1}f(\hat{g}_{t+i})[y_s] \right) \right),
$$
where $\hat{g}_{t+i}$ is the estimated future graph after $i$ steps. The reward for ENS is defined as the margin between two scores:
\begin{equation*}
\label{eq:reward1-ens}
    R_{\pi_S}^{(t)} = e^{-(r_{t+1}-r_{t})}.
\end{equation*}
Similar to the rewards in CRG, if the performance improves after action $a_t$, we obtain $R_{\pi_S}^{(t)}<1$. Additionally, we introduce another reward to penalize the module if the classification performance on the current key propagation graph is unsatisfactory:
\begin{equation*}
\label{eq:reward2-ens}
    \bar{R}_{\pi_S}^{(t)} = 1.5 - f(g_{t+1})[y_s].
\end{equation*}
If the prediction score on class $y_s$ falls below $0.5$, then $\bar{R}_{\pi_S}^{(t)} \geq 1$, which penalizes the ENS module. The final loss of ENS is derived by combining this penalty with the cross-entropy loss of the GNN:
\begin{equation*}
    \mathcal{L}_{\pi_S}^{(t)} = \bar{R}_{\pi_S}^{(t)} \cdot R_{\pi_S}^{(t)}\cdot L_{CE}.
    \label{eq:ens-loss}
\end{equation*}

\header
\textbf{Training Pipeline.}
ENS and CRG modules are interdependent within the KPG framework. On one hand, ENS relies on CRG to generate new responses, ensuring that there are sufficient nodes in the candidate set for selection. On the other hand, CRG requires ENS to provide effective context-response pairs to accurately capture the latent variable distribution of responses. Therefore, we employ an alternative training approach that enables end-to-end learning of both modules.
Specifically, we first fix CRG $\pi_R$ and update ENS $\pi_S$ by minimizing $\mathcal{L}_{\pi_S}$ using Eq.\ \ref{eq:ens-loss} through policy gradient \cite{SuttonMSM99policy}. Next, we fix ENS $\pi_S$ and update CRG $\pi_R$ by minimizing $\mathcal{L}_{\pi_R}$ using Eq.\ \ref{eq:crg-loss}. This iterative process continues until either the early stop condition is met or the maximum number of steps is reached. Algorithm \ref{alg:pipeline} summarizes the learning pipeline of KPG.

To enhance generalization and mitigate over-fitting, we adopt batch training along with an early stop strategy. The final key propagation graphs are used to train a new BiGCN classifier for downstream rumor detection tasks. Following GACL \cite{sun22gacl}, we concatenate the texts of the source post and comments in the propagation graph as the input text for a BERT classifier \cite{Devlin19bert}. The results from two classifiers are fused using the mean operation to derive the final classification results. Additionally, we feed the training events to the model in descending order based on the sizes of their propagation graphs. This strategy ensures that both ENS and CRG can learn from informative propagation patterns at the beginning stage and better handle limited-spread events with small graphs.

\header
\textbf{Time Complexity. }
The time complexity of KPG per epoch is $O(L(Mh+Nh^2))$, where $N$ and $M$ denote the total nodes and edges across all graphs, respectively, $L$ is the maximum number of generation steps, and $h$ is the dimension of hidden layers. The overall cost is primarily determined by the step number $L$ and the cost of the GNN model. In addition, the step number $L$ is a tunable parameter that can be adjusted to balance processing time and model accuracy. Experimental results on the Twitter16 dataset in Table \ref{tab:tau-m} show that KPG achieves superior classification accuracy compared to other baseline methods, even with small $L$ values.

\begin{table*}[t]
\caption{Results on four datasets. NR, FR, TR, UR, and R represents for non-rumor, false rumor, true rumor, unverified rumor, and rumor, respectively. Best and second-best results are highlighted with bold and underlined text. We exclude methods that cannot finish detection within 7 days. The last column reports the average rank over all datasets and metrics of each method. }
\vspace{-3mm}
\label{tab:allres}
\scalebox{0.85}{
\begin{tabular}{lccccccccccccccccc}
\toprule
 & \multicolumn{5}{c}{Twitter15} & \multicolumn{5}{c}{Twitter16} & \multicolumn{3}{c}{Pheme} & \multicolumn{3}{c}{Weibo22} \\  \cmidrule(lr){2-6} \cmidrule(lr){7-11}\cmidrule(lr){12-14}\cmidrule(lr){15-17}
\multirow{-2}{*}{} & Acc. & NR-$F_1$ & FR-$F_1$ & TR-$F_1$ & UR-$F_1$ & Acc. & NR-$F_1$ & FR-$F_1$ & TR-$F_1$ & UR-$F_1$ & Acc. & R-$F_1$ & NR-$F_1$ & Acc. & R-$F_1$ & NR-$F_1$ & \multirow{-2}{*}{Rank}\\\hline
BERT & 0.784 & 0.841 & 0.771 & 0.811 & 0.714 & 0.753 & 0.804 & 0.660 & 0.837 & 0.699 & 0.816 & 0.809 & 0.815 & 0.887 & 0.888 & 0.886 & 12.1\\\hline
RvNN & 0.793 & 0.817 & 0.791 & 0.822 & 0.745 & 0.783 & 0.755 & 0.737 & 0.849 & 0.783 & 0.768 & 0.774 & 0.762 & 0.858 & 0.858 & 0.857 & 12.4 \\
BiGCN & 0.837 & 0.813 & 0.846 & 0.892 & 0.798 & 0.857 & 0.798 & \textbf{0.836} & 0.923 & 0.873 & 0.841 & 0.842 & 0.841 & 0.905 & 0.907 & 0.904 & 6.1\\
EBGCN & 0.851 & 0.827 & 0.864 & 0.891 & {\ul 0.825} & 0.858 & 0.804 & 0.835 & 0.920 & {\ul 0.874} & 0.839 & 0.840 & 0.837 & 0.870 & 0.876 & 0.864& 6.6 \\
RDEA & 0.860 & 0.872 & 0.866 & 0.888 & 0.815 & 0.868 & 0.861 & 0.817 & 0.920 & 0.871 & 0.838 & 0.840 & 0.837 & {\ul 0.918} & {\ul 0.920} & {\ul 0.917} & 5.1 \\ 
GACL & 0.785 & 0.877 & 0.740 & 0.780 & 0.738 & 0.758 & 0.814 & 0.686 & 0.801 & 0.719 & 0.836 & 0.839 & 0.833 & 0.885 & 0.884 & 0.886 & 10.8 \\
TrustRD & {\ul 0.866} & 0.875 & {\ul 0.871} & {\ul 0.896} & 0.821 & 0.869 & 0.844 & 0.827 & {\ul 0.929} & 0.874 & {\ul 0.846} & 0.846 & {\ul 0.846} & {\ul 0.918} & {\ul 0.920} & {\ul 0.917} & 3.1 \\
SMG & 0.828 & 0.861 & 0.840 & 0.869 & 0.736 & 0.841 & 0.812 & 0.812 & 0.898 & 0.839 & 0.830 & 0.830 & 0.830 & - & - & - & 9.4 \\
AdaSNN & 0.798	& 0.763 & 0.767 & 0.844 & 0.772 & 0.792 & 0.711 & 0.783 & 0.871 & 0.800 & 0.820 & 0.811 & 0.827 & - & - & - & 11.8 \\
DCE-RD & 0.827	& 0.850 & 0.817 & 0.855 & 0.776 & 0.826 & 0.808 & 0.785 & 0.884 & 0.821 & 0.834 & 0.833 & 0.835 & - & - & - & 12.5 \\
\hline
GLAN & 0.827 & 0.820 & 0.839 & 0.871 & 0.779 & 0.831 & 0.756 & 0.805 & 0.923 & 0.843 & 0.845 & {\ul 0.849} & 0.840 & 0.902 & 0.903 & 0.902 & 7.6\\
SMAN & 0.853 & {\ul 0.894} & 0.851 & 0.860 & 0.806 & 0.853 & {\ul 0.867} & 0.788 & 0.914 & 0.842 & 0.836 & 0.837 & 0.835 & 0.911 & 0.912 & 0.909 & 6.6 \\
SBAG & 0.862 & 0.876 & 0.862 & 0.887 & 0.821 & {\ul 0.870} & 0.863 & \textbf{0.836} & 0.913 & 0.866 & 0.841 & 0.842 & 0.841 & 0.912 & 0.912 & 0.911 & 4.4 \\ 
\hline
KPG & \textbf{0.893} & \textbf{0.921} & \textbf{{0.898}} & \textbf{0.903} & \textbf{0.847} & \textbf{0.889} & \textbf{0.894} & \textbf{0.836} &  \textbf{0.930} & \textbf{0.896} & \textbf{0.859} & \textbf{0.863} & \textbf{0.854} & \textbf{0.949} & \textbf{0.949} & \textbf{0.948} & 1.0 \\ 
\bottomrule
\end{tabular}
}
\vspace{-3mm}
\end{table*}

\section{Experiment}
\label{sec:experiment}
In this section, we compare our KPG against 12 state-of-the-art competitors on 4 datasets\footnote{Our code is available at \url{https://github.com/kkkkk001/KPG}}. We also conduct early-stage rumor detection and the ablation study to demonstrate the effectiveness of each component of KPG. Finally, we perform the parameter analysis to examine the impact of the parameters on the model performance.

\subsection{Datasets}
\label{sec:subsec-dataset}
We evaluate KPG on three real-world benchmark datasets: Twitter15 \cite{ma17twitter1516}, Twitter16 \cite{ma17twitter1516}, and Pheme \cite{zubiaga16pheme}. Twitter15 and Twitter16 are collected from the Twitter platform and divided into four rumor classes: {\em non-rumor (NR), true rumor (TR), false rumor (FR)}, and {\em unverified rumor (UR)}. The Pheme dataset, also derived from Twitter, is related to five events and is annotated with two labels: {\em rumor (R)} and {\em non-rumor (NR)}. These datasets are commonly used in existing research studies \cite{ma18rvnn,bian2020bigcn,sun22gacl} on rumor detection.

The Weibo dataset used in \cite{bian2020bigcn, Ma16rnn} lacks critical information required for several baselines, including original comment text and author details. However, these details cannot be fully recovered due to the presence of many deleted user accounts and posts on Weibo. In addition, the propagation patterns and topics of rumor in real-world social networks have evolved rapidly. To address these issues, we have constructed a new dataset \textbf{Weibo22}, which is also available in our code repository.
Events in Weibo22 are collected from Sina Weibo, one of the largest social media platforms in China. The dataset covers events from November 2019 to March 2022, with more than half of the events related to the COVID-19 pandemic. Specifically, the events in Weibo22 are divided into two categories, rumor and non-rumor, based on information provided by Weibo Community Management Center \cite{weibocommunity} and China Internet Joint Rumor Debunking Platform \cite{piyao}. By using the Weibo API \cite{weiboapi}, we collected 4,174 source posts with 960,000 microblogs, including reposts and comments in their propagation graphs. For each microblog, we also collected user profile information, including the number of followers, number of friends, verification status, verification type, and verification reason. 
Detailed statistics of these four datasets are shown in Table \ref{tab:dataset}.

\subsection{Experimental Settings}
\label{sec:subsec-exp-setting}
\textbf{Baselines. }
We compare KPG with twelve state-of-the-art baselines, including BERT, RvNN, BiGCN, EBGCN, RDEA, GACL, TrustRD, SMG, AdaSNN, DCE-RD, GLAN, SMAN, and SBAG.
Detailed descriptions of these baselines can be found in Appendix \ref{sec:dataset-baseline}. 

\header
\textbf{Metrics and evaluation protocol. }
We adopt Accuracy (Acc.) and micro $F_1$ score ($F_1$) for each class as our evaluation metrics. Notice that several baselines \cite{bian2020bigcn, wei2021ebgcn, he21rdea, sun22gacl, liu2023trustrd} utilize a batch-wise averaging approach. 
However, this may introduce performance variance, particularly when the final batch is smaller than the others. For example, consider a dataset with 110 records and a batch size of 100. If the accuracy for the first batch is 0.8 and for the second batch is 0.9, the batch-wise averaging yields a result of 0.85, obviously skewed by the smaller second batch. To ensure a reliable evaluation, we calculate the average accuracy across all batches in each epoch, mitigating any impact from variations in the final batch size.

\subsection{Experimental Results}
\label{sec:subsec-exp-results}
Table~\ref{tab:allres} shows the performance of each model on four real-world datasets. Note that we exclude any baseline that cannot complete detection within 7 days. As observed, our KPG consistently achieves the best accuracy across all datasets, demonstrating the effectiveness and generalization capability of our proposed RL-based key propagation graph generation method. 
On the Twitter15 and Twitter16 datasets, our KPG outperforms the second-best results by achieving an increase of 2.7\% in the $F_1$ score for the non-rumor class. Additionally, on the Pheme and Weibo22 datasets, KPG leads by 1.4\% and 2.9\% in the $F_1$ score for the rumor class compared to the second-best performances.

Moreover, KPG achieves the highest average ranking, highlighting its superior ability to generalize across diverse datasets and metrics. Specifically, compared to TrustRD, the model with the second highest average rank, KPG takes the lead by 2.7\% and 2\% in terms of accuracy on Twitter15 and Twitter16 datasets, respectively. Furthermore, KPG outperforms models that use additional user information, including GLAN, SMAN, and SBAG. Compared to these three models, KPG leads by up to 6.6$\%$ on Twitter15 and 5.8$\%$ on Twitter16 in terms of accuracy, further showcasing the effectiveness of KPG.  

On our newly collected dataset, Weibo22, KPG outperforms all competitors across all metrics. Compared to the second-best baseline, KPG takes a lead by 3.1\%, 2.9\%, and 3.1\% in terms of accuracy, $F_1$ score for rumor, and $F_1$ score for non-rumor, respectively, This again demonstrates that our key graph generation method effectively reduces noise while extracting more useful information.

\begin{figure}[t]
\centering
\vspace{-1mm}
\begin{small}
\begin{tabular}{cc}
    \multicolumn{2}{c}{\includegraphics[width=0.95\linewidth]{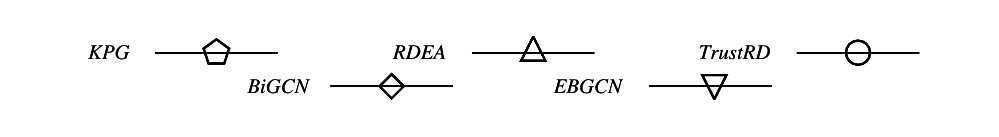}}  \\[-3mm]
    \hspace{-4mm}\includegraphics[height=30mm]{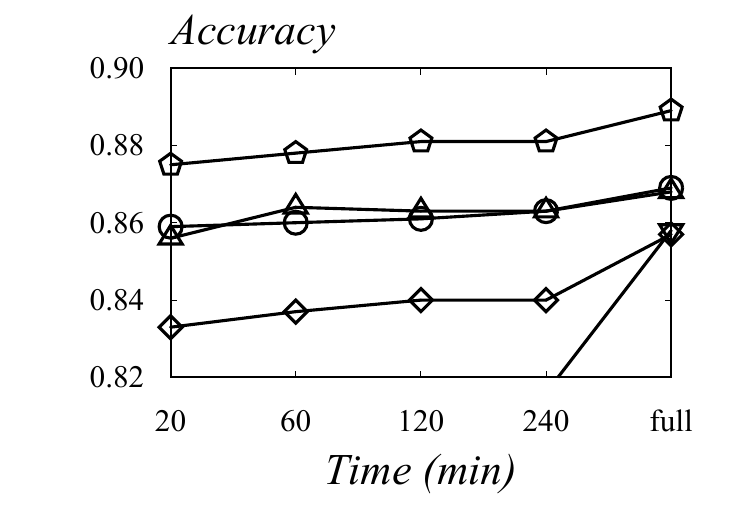} &
    \includegraphics[height=30mm]{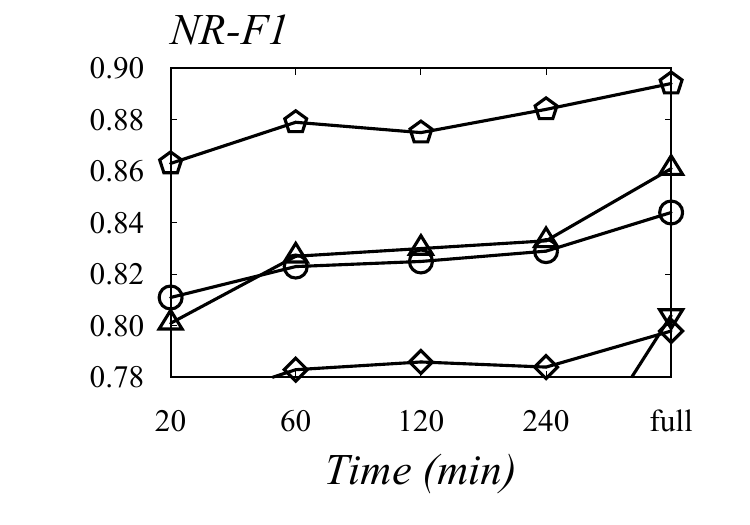} \\
    [-1mm]
    (a) Accuracy & (b) F1 score on NR \\[-1mm]
\end{tabular}
 \vspace{-3mm}
\caption{Early stage rumor detection on Twitter16.}
\label{fig:early_acc_f1}
\end{small}
\vspace{-4mm}
\end{figure}
\subsection{Early Stage Rumor Detection}
We also conduct experiments to validate the performance of KPG in detecting rumors during the early stage of their spread. To achieve this, we set a temporal threshold $\Delta$ to filter nodes within each propagation graph. Specifically, for every node in a propagation graph, we compute the time difference between the publication time of the comment (or retweet) and the creation time of the root claim. Nodes with a time difference less than $\Delta$ are retained, while those exceeding $\Delta$ are excluded. We conduct experiments with varying $\Delta \in \{20, 60, 120, 240\}$ minutes. We compare KPG with the top four baselines in terms of average rank, excluding methods that require additional author information. The results for TrustRD, RDEA, BiGCN, EBGCN, and KPG in terms of accuracy and $F_1$ score on non-rumor events are presented in Figure \ref{fig:early_acc_f1}. The results for $F_1$ scores on the other three classes are similar to those of the NR class and thus are omitted for brevity.

As we can see, during the early stage of spread, the textual features and topology information contained in the propagation graphs diminish, leading to a general decrease performance. However, the performance of our KPG consistently surpasses that of other baselines across varying $\Delta$ values. When $\Delta=20$, KPG outperforms TrustRD, the second-best baseline, in both accuracy and $F_1$ score for the non-rumor class, even though TrustRD utilizes full propagation graphs. 
This superior performance further demonstrates the effectiveness of our KPG, particularly the CRG module, which generates realistic and informative responses and plays a crucial role in early-stage rumor detection.

\subsection{Ablation Study and Parameter Analysis}
In this section, we first examine the effectiveness of each sub-module in KPG, and then analyze the effects of hyperparameters. 

\begin{table}[t]
\caption{Ablation study on Twitter15. }
\vspace{-3mm}
\label{tab:ablat15}
\centering
\begin{tabular}{cccccc}
\toprule
 & {Acc.} & {NR-$F_1$} & {FR-$F_1$} & {TR-$F_1$} & {UR-$F_1$}\\ 
 \midrule
KPG & \textbf{0.893} & \textbf{0.921} & \textbf{0.898} & \textbf{0.903} & \textbf{0.847} \\ 
KPG$\backslash$ens & 0.886 & 0.920 & 0.877 & 0.900 & 0.847 \\ 
KPG$\backslash$crg & 0.880 & 0.916 & 0.869 & 0.892 & 0.843 \\ 
KPG$\backslash$reward & 0.883 & 0.915 & 0.875 & 0.896 & 0.845 \\ 
\bottomrule
\end{tabular}
\end{table}

\begin{table}[t]
\vspace{-4mm}
\caption{Ablation study on Twitter16. }
\vspace{-3mm}
\label{tab:abla}
\centering
\begin{tabular}{cccccc}
\toprule
 & {Acc.} & {NR-$F_1$} & {FR-$F_1$} & {TR-$F_1$} & {UR-$F_1$}\\ 
 \midrule
KPG & \textbf{0.889} & \textbf{0.894} & \textbf{0.836} & \textbf{0.930} & \textbf{0.896} \\ 
KPG$\backslash$ens & 0.865 & 0.882 & 0.810 & 0.910 & 0.855 \\ 
KPG$\backslash$crg & 0.861 & 0.878 & 0.806 & 0.908 & 0.849 \\ 
KPG$\backslash$reward & 0.865 & 0.880 & 0.808 & 0.914 & 0.855 \\ 
\bottomrule
\end{tabular}
\vspace{-4mm}
\end{table}

\header
\textbf{Ablation Study.}
We implement three variants of KPG: 
\begin{itemize}[topsep=0.5mm, partopsep=0pt, itemsep=0pt, leftmargin=10pt]
    \item KPG$\backslash$ens: it randomly selects out-neighbors of existing nodes;
    \item KPG$\backslash$crg: it removes the CRG module and stops generating extra candidate nodes for small propagation graphs; 
    \item KPG$\backslash$reward: it eliminates rewards during training.
\end{itemize}
Tables \ref{tab:ablat15}-\ref{tab:abla} report the results of KPG and its three variants on the Twitter15 and Twitter16 datasets, respectively. As we can observe, the absence of any component leads to a decrease in rumor detection performance. This decline not only demonstrates the effectiveness of each module but also underscores the importance of synergy of all three components.

\begin{table}[t]
\caption{Results with varying $\epsilon$ on Twitter16.}
\vspace{-3mm}
\label{tab:eps}
\centering
\begin{tabular}{ccccccc}
\toprule
$\epsilon$ & 1.0 & 0.8 & 0.6 & 0.4 & 0.2 & 0.0 \\
\midrule
Acc. & 0.880 & 0.889 & 0.880 & 0.878 & 0.880 & 0.880 \\
NR-$F_1$ & 0.884 & 0.894 & 0.881 & 0.884 & 0.880 & 0.888 \\
FR-$F_1$ & 0.831 & 0.836 & 0.827 & 0.824 & 0.831 & 0.827 \\
TR-$F_1$ & 0.922 & 0.930 & 0.927 & 0.922 & 0.924 & 0.923 \\
UR-$F_1$ & 0.884 & 0.896 & 0.885 & 0.882 & 0.884 & 0.883 \\
\bottomrule
\end{tabular}
\end{table}
\begin{table}[t]
\vspace{-3mm}
\caption{Results with varying $\tau$ and $l$ on Twitter16. }
\vspace{-3mm}
\centering
\label{tab:tau-m}
\scalebox{1.0}{
\begin{tabular}{cccccccc}
\toprule
\multirow{2}{*}{} & \multicolumn{4}{c}{$\tau$} & \multicolumn{3}{c}{$l$} \\
\cmidrule(lr){2-5} \cmidrule(lr){6-8}
 & 0 & $2^1$ & $2^2$ & $2^3$ & 0 & 5 & 10 \\
 \midrule
\multicolumn{1}{c}{Acc.} & 0.832 & 0.875 & 0.883 & 0.889 & 0.879 & 0.872 & 0.889  \\
\multicolumn{1}{c}{NR-$F_1$} & 0.763 & 0.878 & 0.896 & 0.896 & 0.891 & 0.870 & 0.894\\
FR-$F_1$ & 0.819 & 0.825 & 0.824 & 0.836 & 0.822 & 0.823 & 0.836  \\
TR-$F_1$ & 0.887 & 0.923 & 0.932 & 0.930 & 0.919 & 0.918 & 0.930 \\ 
UR-$F_1$ & 0.859 & 0.872 & 0.880 & 0.896 & 0.882 & 0.875 & 0.896 \\
\bottomrule
\end{tabular}
}
\vspace{-3mm}
\end{table}

\header
\textbf{The Effect of Parameter $\epsilon$ in Candidate Selection.}  
The parameter $\epsilon \in [0, 1]$ serves as a trade-off between local and global candidates. A larger $\epsilon$ indicates a higher probability for ENS to select the next node from the local neighbors of currently selected nodes, while a smaller $\epsilon$ increases the probability of selecting from nodes that have not yet been selected. We vary the parameter $\epsilon$ from 0 to 1 and report the corresponding results on Twitter16 in Table \ref{tab:eps}. The results on other datasets are similar to those on Twitter16 and thus are omitted.
In general, the model performance is influenced by the balance between local and global exploration. Specifically, on Twitter16, optimal performance is achieved when $\epsilon=0.8$. A similar performance trend is observed across other datasets. Therefore, we set $\epsilon=0.8$ for KPG in our experiments.

\header
\textbf{The Effect of the Maximum Generation Steps.} 
Table \ref{tab:tau-m} reports the experimental results on Twitter16 with varying $\tau$, which controls the maximum size of the generated key propagation graph. We set the maximum generation steps to $\tau$ times the median size of original graphs, where $\tau$ is chosen from $\{0, 2^1, 2^2, 2^3\}$. When $\tau=0$, only the root node is used for classification, resulting in the worst performance across all metrics. This highlights the importance of leveraging propagation graphs for effective rumor detection. As $\tau$ increases, we observe a general performance improvement across all metrics. When $\tau=2^3$, KPG achieves peak performance on Twitter16. This can be attributed to the relatively small average graph size in the Twitter16 dataset. The results of the ablation study further support this observation. Without the CRG component, reaching the maximum generation step becomes challenging, resulting in the most significant performance decline among three variants. Compared to other baselines,  KPG achieves superior accuracy across various $\tau$ values, even when $\tau$ is relatively small.

\header
\textbf{The Effect of the Maximum Step $l$ in Modified Rollout.} Table \ref{tab:tau-m} also reports the performance on Twitter16 with varying $l$, which is the maximum steps of the Rollout used to derive rewards for ENS. We conduct experiments with $l \in \{0, 5, 10\}$. As we can observe, the model performance peaks at  $l=10$, which demonstrates the effectiveness of evaluating each node from a long-term perspective.

\begin{figure}[t]
    \centering
    \begin{small}
    \includegraphics[width=0.45\textwidth]{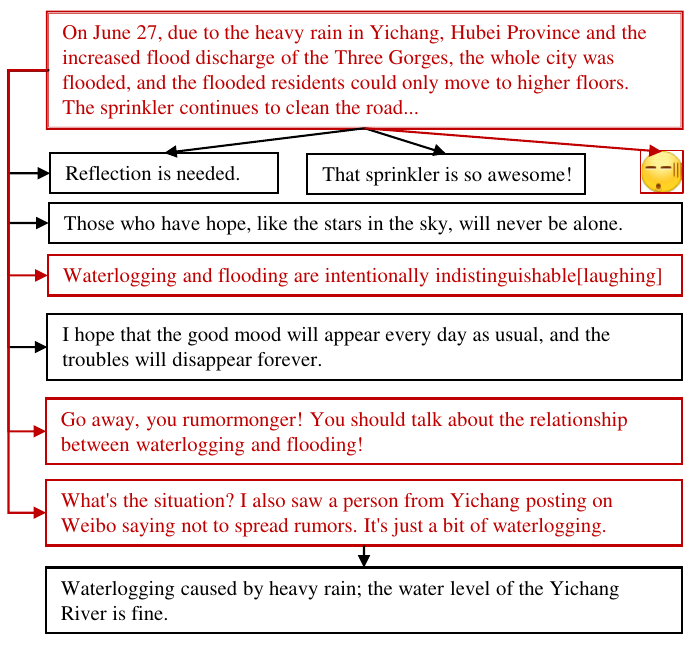}
    \vspace{-3mm}
    \caption{An example of the generated propagation graph. }
    \label{fig:large}
 \vspace{-4mm}
    \end{small}
\end{figure}

\subsection{Case Study}
To illustrate the functionality of our KPG, we present the key propagation graph of a widespread rumor \cite{rumorcase} from the Weibo22 dataset in Figure \ref{fig:large}. 
In the original propagation graph, the event cannot be correctly classified by BiGCN. However, with the key propagation graph generated by KPG, the rumor can be correctly identified. 
Our observations indicate that KPG selects nodes (highlighted in red) that enhance the discriminative capability of rumor detection while disregarding irrelevant and noisy nodes. In particular, strong emotional responses are selected to refute the rumor.

\section{Conclusion}
\label{sec:conclustion}
In this paper, we propose KPG, a reinforcement learning-based model for rumor detection, comprised of two components: ENS and CRG. With the cooperation of two modules, KPG effectively identifies indicative substructures for events with noisy propagation information and generates realistic, reliable, and informative responses for events with insufficient propagation. 
The two components are trained alternately in an end-to-end framework, guided by our carefully designed rewards, to improve the discriminative ability of the generated key propagation graphs. 
Extensive experiments conducted on four real-world datasets demonstrate that KPG achieves state-of-the-art performance. 
Additionally, we construct a new dataset, Weibo22, which contains posts with more recent dates and topics, potentially contributing to the development of rumor detection-related research.

\begin{acks}
This work is supported by the RGC GRF grant (No. 14217322), Hong Kong ITC ITF grant (No. MRP/071/20X), and Tencent Rhino-Bird Focused Research Grant. 
\end{acks}

\bibliographystyle{ACM-Reference-Format}
\bibliography{ref}

\appendix

\section{Notations}
\label{sec:notation}
We summarize the frequently used notations in Table \ref{tab:notation}. 

\section{Additional Related Work}
\label{sec:related-work2}

\header
\textbf{Heterogeneous Graph-based Fake News Detection.} Instead of propagation graphs reflecting the rumor-spreading patterns, another line of research \cite{yuan19glan, yuan20sman, zhen22sbag} constructs heterogeneous graphs of publishers, tweets, and users to integrate both local and global relationships. Based on these heterogeneous graphs, a series of models have been proposed. For instance, GCAN \cite{lu20gcan} uses a dual co-attention mechanism to provide reasonable explanations by learning features from four aspects. Cui et al. \cite{cui22hetscan} propose encoding metapaths from heterogeneous graphs. FinerFact \cite{jin22finerfact} reasons for important evidence from a constructed claim-evidence graph using a mutual reinforcement mechanism \cite{mutualrein}. SureFact \cite{yang22surefact} further leverages RL to measure the importance of nodes and conducts subgraph reasoning. Additionally, DECOR~\cite{wu2023decor} and PSGT~\cite{zhu2024propagation} incorporate auxiliary information to enhance the detection of fake news, such as degrees of news articles and malicious users during the user-article engagement.
Note that additional information used to construct the heterogeneous graphs, such as news-user discussion graphs, user-user interaction graphs, news-post similarity graphs, is not included in the datasets we used. To ensure a fair comparison, we exclude these methods from our experiments.

\header
\textbf{Graph Generation.}
Graph generation is the task of learning the distribution of the observed graphs to generate new realistic graphs. One of the criteria used for classifying graph generative models is the generation process, including one-shot and sequential \cite{guo23gensurv}. GraphRNN \cite{you18grnn} lays the foundation for subsequent node-by-node deep auto-regressive models. After that, MolecularRNN \cite{popova2019molecularrnn} extends GraphRNN to generate realistic molecular graphs with optimized properties. DeepGMG \cite{li18dgmg} defines a sequential graph generation process as a sequence of decisions. Faez et al. summarize in \cite{FaezOBR21} that current reinforcement learning-based graph generators typically use a sequential generation strategy. GCPN \cite{you2018gcpn} and its related methods \cite{ShiY019,karimi20,khemchandani20,trivedi20gopt} employ a step-by-step approach in generating molecular graphs by formulating the problem as a Markov decision process. 
GraphOpt \cite{trivedi20gopt} simulates the decision process for graph construction and searches for the reward function by optimizing the objective of assigning scores to the observed graphs. These graph generators mainly focus on the generation of chemical molecules, while we design a new dynamically updated candidate set and effectively combine node features with structural information for generating propagation graphs on social media.

\header
\textbf{Response Generation.}
The generation of responses for dialogue services is an important task in natural language processing. Earlier attempts \cite{serban17vhred}, integrate the concept of variational auto-encoders (VAE) \cite{kingma14vae,rezende14vae} into response generation. After that, advanced studies \cite{ShenSND18,wu20psncvae,zhao17kgcvae,sun20sepacvae,ijcai2022pcvae} are proposed to enhance response diversity based on CVAE \cite{Sohn2015cvae}. For instance, (kg)CVAE \cite{zhao17kgcvae} integrates linguistic prior knowledge. SepaCVAE \cite{sun20sepacvae} introduces contrastive learning to leverage group information. PAGenerator \cite{wu20psncvae} adds regularization terms to guide the generation of personal-aware and relevant responses. 
Instead of relying solely on news content \cite{qian18rps}, we exploit propagation graphs to learn the latent variable distribution of responses that are more helpful in identifying rumors in social media.

\section{Algorithm}
The complete learning pipeline is presented in Algorithm~\ref{alg:pipeline}.

\begin{table}[t]
\caption{Frequently used notations.}
\vspace{-3mm}
\label{tab:notation}
\centering
\begin{tabular}{|l|l|}
\hline
Notations & Descriptions \\
\hline
$r, s, y_s$ & Source post, event, and its ground-truth label. \\
\hline
$G,V,E,\vect{X}$ & Input propagation graph, node set, edge set,\\ 
& and feature matrix. \\
\hline
$g_t, G_t$ & Key propagation graph and candidate \\
& graph at step $t$.  \\
\hline
$v_t, e_t$ & The newly added node and edge at step $t$. \\
\hline
$C_{t+1}$ & Candidate set for ENS to select $v_t$.\\
\hline
$\epsilon, \gamma$ & Trade-off parameter and threshold of \\ & candidate sets. \\
\hline
$l$ & The max step in modified Rollout.\\
\hline
$\tau$ & Controlling the max generation step.\\
\hline
\end{tabular}
\end{table}

\section{Datasets and Baselines}
\label{sec:dataset-baseline}

\header
\textbf{Datasets.}
In our experiment, we utilize three commonly used datasets, Twitter15, Twitter16, and Pheme, to evaluate the effectiveness of our KPG. Besides, to explore the rumors with more recent dates and topics, we have collected a new dataset, Weibo22. The statistics of these four datasets are presented in Table \ref{tab:dataset}. Due to deleted tweets and deactivated accounts, our dataset statistics slightly deviate from those originally reported in~\cite{ma17twitter1516}.

\header
\textbf{Baselines.}
Based on code availability and task consistency, we selected twelve state-of-the-art baseline models. The included competitors are introduced below.

\begin{algorithm}[t]
\DontPrintSemicolon
\SetKwFor{For}{for}{do}{} 
\SetKwInOut{Input}{Input}\SetKwInOut{Output}{Output}
    \Input{Set of events $\{s_0, s_1, \cdots\}$, pre-trained BiGCN classifier $f(\cdot)$, maximum step $L$}
    \Output{Key propagation graph for each event,  $\pi_S$, $\pi_R$}
    Initialize $\pi_S$, $\pi_R$ \;
    \For{each epoch}{
        \For{each batch}{
            \For{step $t=0$ to $L$}{
                Train $\pi_R$ using Eq. \ref{eq:crg-loss} with $\pi_S$ fixed \;
                ${G}_{t+1} \leftarrow \pi_R({G}_t)$ \;
                $\vect{p}_t \leftarrow \pi_S({G}_{t+1}, g_t)$  \;
                Sample action $a_t=(v_t, e(v_t))$ based on $\vect{p}_t$ \;
                Transfer to state $g_{t+1}=(V_{g_t} \cup v_t, E_{g_t} \cup e(v_t), \vect{X}_{g_t}\cup {\vect{X}}[v_t])$ \;
                Train $\pi_S$ using Eq. \ref{eq:ens-loss} with $\pi_R$ fixed \;
                
            }
        }
    }
\caption{Training Pipeline of KPG}
\label{alg:pipeline}
\end{algorithm} 

\begin{itemize}[topsep=0.5mm, partopsep=0pt, itemsep=0pt, leftmargin=10pt]
    \item BERT \cite{Devlin19bert}: a powerful pre-trained language model based on bidirectional transformers;
    \item RvNN \cite{ma18rvnn}: it employs tree-structured RNNs with GRUs to extract representations from the propagation structure in bottom-up and top-down manners; 
    \item BiGCN \cite{bian2020bigcn}: a GCN-based model that learns features from both propagation and dispersion structures; 
    \item EBGCN \cite{wei2021ebgcn}: it uses a Bayesian approach to adjust the weights of uncertain relations and enforces consistency on latent relations using an edge-wise training framework; 
    \item RDEA~\cite{he21rdea}: a contrastive self-supervised learning-based method with event augmentation; 
    \item GACL~\cite{sun22gacl}: it designs graph perturbation methods based on adversarial and supervised contrastive learning;
    \item TrustRD ~\cite{liu2023trustrd}: it incorporates self-supervised learning and a Bayesian network to derive trustworthy predictions;
    \item SMG~\cite{yang21smg}: it learns graph representations from a sequence of subgraphs to better capture task-relevant substructures and skip noisy parts; 
    \item AdaSNN~\cite{li23adasnn}: it generates critical subgraphs with a bi-level mutual information enhancement mechanism optimized in the reinforcement learning framework;
    \item DCE-RD~\cite{zhang23dcerd}: it exploits diverse counterfactual evidence to serve as multi-view interpretations on event graphs;
    \item GLAN \cite{yuan19glan}: it learns local semantic and global structural information via attention mechanisms on the heterogeneous graph;
    \item SMAN~\cite{yuan20sman}: it adopts a structure-aware multi-head attention module on the heterogeneous graph to optimize the user credibility prediction and rumor detection task jointly; 
    \item SBAG~\cite{zhen22sbag}, it adopts a pre-trained MLP to capture social bot features in the propagation graph and uses it as a scorer to train a social bot-aware GNN for rumor detection.
\end{itemize}

\header{\bf Remark.} 
SBAG requires additional datasets to train the social bot detection model while such additional information is unavailable in our experiments. Thus, the second-best performing variant reported in the original paper, SBAG-s, which scores the possibility of bots randomly, is compared in our experiment. For the BERT baseline, following previous settings~\cite{sun22gacl}, we concatenate the source post and all comment posts in the same event as the input texts to fine-tune a BERT-based classifier.

\begin{table}[t]
\caption{Statistic of datasets.  }
\vspace{-3mm}
\label{tab:dataset}
\centering
\begin{tabular}{ccccc}
\toprule
Statistic & Twitter15 & Twitter16 & Pheme & Weibo22 \\  
\midrule
\# posts & 41,194 & 18,618 & 67,238 & 961,962 \\  
\# users & 34,668 & 16,805 & 34,287 & 705,831 \\  
\# events & 1,490 & 818 & 3,720 & 4,174 \\  
\# Non-rumors & 374 & 205 & 1,860 & 2,087 \\  
\# False rumors & 370 & 205 & 1,860 & 2,087 \\  
\# True rumors & 372 & 205 & 0 & 0 \\  
\# Unverified rumors & 374 & 203 & 0 & 0 \\  
Avg. \# posts / event & 28 & 23 & 18 & 230 \\  
Med. \# posts / event & 16 & 13 & 15 & 7 \\  
Max \# posts / event & 304 & 250 & 346 & 30,791 \\  
Min \# posts / event & 1 & 1 & 2 & 1 \\  
\bottomrule
\end{tabular}
\end{table}

Among the baseline models, BERT is the only one that utilizes text features without propagation structures. RvNN, BiGCN, EBGCN, RDEA, GACL, and TrustRD are rumor detection methods utilizing the propagation structures of social network posts. Additionally, we also include SMG and AdaSNN, which are not specifically tailored for rumor propagation graphs, but for a general spectrum of graphs. These two graphs are proposed with a focus on handling graphs containing noisy or unreliable information. Moreover, GLAN, SMAN, and SBAG employ additional user information to construct heterogeneous graphs, while other competitors, along with our KPG, do not use any user information, but only the propagation structures in rumor detection. 

\section{Parameter Setting}
\label{sec:params}

Following \cite{bian2020bigcn, wei2021ebgcn, he21rdea, sun22gacl}, we randomly split the datasets into five parts and conduct 5-fold cross-validation to obtain the final results.  For each dataset, we use the same data split on all methods for fair comparison. 
We set the maximum step of the key propagation graph generation equal to $\tau$ times the median size of the original graphs in each dataset, and we utilize grid search to set $\tau$ from $\{2^1, 2^2, 2^3\}$. For Weibo22, considering that the median size of the original graphs is much smaller than the average size, we add an additional choice, $s_{avg}$, to the search list, where $s_{avg}$ is the average size of the original graphs. 
We set the threshold of candidates $\gamma=5$ in CRG. We set the maximum step of rollout in ENS to $10$, and the trade-off parameter between two candidate sets $\epsilon=0.8$. We pre-train the BiGCN classifier for 30 epochs to derive rewards.

After the generation of key propagation graphs, we train another BiGCN classifier on the key graphs with 200 epochs to obtain the final classification results. Our KPG is optimized by Adam algorithm \cite{Kingma14adam}. The learning rate is initialized to $5\times10^{-4}$ and gradually decreases during training with a decay rate of $10^{-4}$. The dimension of hidden feature vectors in all modules is set to 64, and the batch size is set to 128.

\end{document}